\documentclass[aip,pof,twocolumn,showpacs,groupedaddress]{revtex4}  
\usepackage{graphicx}  
\usepackage{dcolumn}   
\usepackage{bm}        
\usepackage{amssymb}   
\usepackage{ulem}
\usepackage{epsfig}
\usepackage{array}
\usepackage{amsmath}
\usepackage{lscape}
\usepackage{color}
\usepackage{graphicx,subfigure}

\newcommand{\re}[1]{~(\ref{#1})}
\newcommand{\beq}{\begin{equation}}
\newcommand{\eeq}{\end{equation}}
\newcommand{\beqn}{\begin{eqnarray}}
\newcommand{\eeqn}{\end{eqnarray}}
\newcommand{\la}{\lambda}
\newcommand{\ld}{\lambda}
\begin{document}
\title{Dissipation scales of kinetic helicities in turbulence}
\author{T. Lessinnes,$^{1}$ F. Plunian,$^2$ R. Stepanov$^3$ and D. Carati,$^1$}

\email{Thomas.Lessinnes@ulb.ac.be, Franck.Plunian@ujf-grenoble.fr, rodion@icmm.ru, dcarati@ulb.ac.be}

\affiliation{$^1$ Physique Statistique et Plasmas, CP231, Facult\'e des Sciences,
Universit\'e Libre de Bruxelles, B-1050 Bruxelles, Belgium;\\
 $^2$ Universit\'e Joseph
  Fourier, CNRS, Laboratoire de G\'eophysique Interne et de
  Tectonophysique, 38041 Grenoble, France\\
$^3$ Institute of Continuous Media Mechanics of the Russian Academy of Science, 614013 Perm, Russia} \date{\today}
\begin{abstract}
A systematic study of the influence of the viscous effect on both the spectra and the nonlinear fluxes of conserved as well as non conserved quantities in Navier-Stokes turbulence is proposed. This analysis is used to estimate the helicity dissipation scale which is shown to coincide with the energy dissipation scale. However, it is shown using the decomposition of helicity into eigen modes of the curl operator, that viscous effects have to be taken into account for wave vector smaller than the Kolomogorov wave number in the evolution of these eigen components of the helicity.
\end{abstract}
\maketitle

\section{Introduction}

In two recent papers it was suggested that dissipation of kinetic helicity occurs at a scale $k_H^{-1}$ larger than the Kolmogorov scale $k_E^{-1}$. This was justified on dimensional grounds \cite{Ditlevsen01a} as well as using a GOY shell model of turbulence \cite{Ditlevsen01b}. In contrast using a different shell model of turbulence based on helical wave decomposition, both scales were found to be equal $k_H=k_E$ \cite{Chen03}. In addition, direct numerical simulations, also presented in \cite{Chen03}, seem to confirm the latter result though, as noted by the authors, the computational limitations prevent to have a Reynolds number sufficiently large to really discriminate between both scenarii.

The purposes of the present work are to investigate further the possible existence of a specific helicity dissipation scale and to understand why two shell models do exhibit different helicity behaviours while their energy spectrum are very much similar. Part of this apparent contradiction comes from the very definition of the dissipation scale. Indeed, in the Kolmogorov theory, there is no ambiguity. The scale at which the energy dissipation terms are no longer negligible when compared to the non-linear fluxes of energy corresponds to the scale at which the energy spectrum departs from the Kolmogorov power law. This scale marks the end of the cascade process as well as the beginning of energy spectrum fall off.

The situation is less clear for non conserved quantities such has the positive $H^+$ and the negative helicity $H^-$ defined respectively as the helicity carried on by the eigenvectors of the curl operator with positive and negative eigenvalues. Generally, for non conserved quantities $Q$, we propose to refer to the dissipation scale as the scale after which the dissipative term \textit{dominates} the dynamics so that the spectrum of $Q$ falls off. Such a scale might very well differ from the scale, referred hereafter as the \textit{viscous scale}, at which the dissipative terms \textit{start} to play a role in the dynamics of~$Q$. Indeed, for a non-conserved quantity, the nonlinear term might very well compensate for the increase of dissipation in part of the high wave number range after the viscous scale and prevent the spectrum to fall off even if dissipation is active. In general, the viscous scale should be smaller than the dissipation scale. However, for conserved quantities, both the viscous and the dissipation scales coincide.

A general discussion on the determination of dissipation scale is presented in Section II for conserved as well as non conserved quantities. The specific case of the two conserved quantities in three-dimensional turbulence, the energy and the helicity, is discussed in Section III. The positive and negative helicities, which are not conserved quantities, are discussed in Section IV. Shell models describing the high Reynolds number behaviour of turbulence are discussed in Section V. Both models used in references~(\cite{Ditlevsen01b}) and~(\cite{Chen03}) are introduced and analyzed numerically in Section IV. It is shown very clearly that the dissipation scales for the helicity and the energy coincide and are given by the Kolmorogov length scale. Moreover, the dissipation scale for the positive and negative helicities also corresponds to the energy dissipation scale. However, the analysis of their fluxes allows to identify very clearly a viscous scale for both $H^+$ and $H^-$ that is smaller than the energy dissipation scales.

\section{Dissipation scales in turbulent systems with cascades}

Before discussing the specific problem of energy or helicity dissipation scale, we consider a general quadratic quantity $Q$ that is not necessarily conserved by the non-linearities of the Navier-Stokes equation:
\begin{equation}
Q=\int_V d^3{\bf r}\ a({\bf r}) b({\bf r}) = \int d^3{\bf k}\ \tilde{a}({\bf k})\tilde{b}({\bf k})^*  +\textrm{c.c.}\,.
\end{equation}
Here, $a({\bf r})$ and $b({\bf r})$ are two fields and $\tilde{a}({\bf k})$ and $\tilde{b}({\bf k})$ are their Fourier transforms. In the following, the system is assumed to be statistically isotropic. In that case, it is convenient to introduce the spectrum $Q(p)$, so that:
\begin{equation}
Q=\int dp\ Q(p).
\end{equation}
The parts of this quantity that are represented by modes such that $|{\bf k}|<\kappa$ and $|{\bf k}|>\kappa$ are denoted respectively by $Q^<(\kappa)$ and $Q^>(\kappa)$:
\begin{align}
Q^<(\kappa)&=\int_{|{\bf k}|<\kappa}\ d^3{\bf k}\ \ \tilde{a}({\bf k})\tilde{b}({\bf k})^*  +\textrm{c.c.},\\
Q^>(\kappa)&=\int_{|{\bf k}|>\kappa}\ d^3{\bf k}\ \ \tilde{a}({\bf k})\tilde{b}({\bf k})^*  +\textrm{c.c.}\, , \\
& Q= Q^<(\kappa) +Q^>(\kappa) \quad \forall \kappa.
\end{align}
Their evolution is given by:
\begin{subequations}
\label{general_balance}
\begin{eqnarray}
\partial_t Q^<(\kappa)&=&s_Q-\Pi_Q^<(\kappa)-d_Q^<(\kappa) \label{GB1}\\
\partial_t Q^>(\kappa)&=&-\Pi_Q^>(\kappa)-d_Q^>(\kappa) \label{GB2}
\end{eqnarray}
\end{subequations}
where $s_Q$ is the source of $Q$ here injected by a forcing process in the largest scales of the system ($k_F$) so that $k_F<\kappa$. In that case, the source term is independent of $\kappa$. The nonlinearity contributions to the evolution of $Q^<$ and $Q^>$ are noted respectively $\Pi_Q^<(\kappa)$ and $\Pi_Q^>(\kappa)$. They correspond to fluxes respectively outward and inward the sphere of radius $\kappa$ if $Q$ is a conserved quantity. The dissipation of $Q$ in the modes $|{\bf k}|<\kappa$  (resp. $|{\bf k}|>\kappa$) is noted $d_Q^<(\kappa)$ (resp. $d_Q^>(\kappa)$).

In the following, the dissipative processes are assumed to be represented by viscous type terms, so that:
\begin{subequations}\label{dissipation}
\begin{eqnarray}
d^<(\kappa)_Q&=&2 \nu \int_0^\kappa \ dp\, p^2\, Q(p)\label{diss_small}\\
d^>(\kappa)_Q&=&2 \nu \int_\kappa^\infty \ dp\, p^2\, Q(p).\label{diss_large}
\end{eqnarray}
\end{subequations}

If the system undergoes a cascading process that transfers $Q$ from the forcing scales to small scales, the non-linear transfer at scale $\kappa$ should be characterised by a typical time scale that will be denoted $\tau^\textrm{nl}_Q(\kappa)$. On the other hand, dissipation processes should also be characterized by a time scales $\tau^\textrm{diss}_Q(\kappa)$. In the case of viscous type dissipation, $\tau^\textrm{diss}_Q(\kappa)=1/(\nu\, \kappa^2)$. The comparison of these characteristic time scales can be used to estimate the end of the cascade range (usually referred to as the inertial range as long as kinetic energy is concerned). Indeed, in the range dominated by the non-linear interactions, $\tau^\textrm{nl}_Q(\kappa)<\tau^\textrm{diss}_Q(\kappa)$ since non-linear interactions should be faster than dissipative processes. On the contrary, in the dissipation range $\tau^\textrm{nl}_Q(\kappa)>\tau^\textrm{diss}_Q(\kappa)$. An estimate of the dissipation scale $k^D_Q$ is thus:
\begin{equation}
\tau^\textrm{nl}_Q(k^D_Q) \approx \tau^\textrm{diss}_Q(k^D_Q)
\end{equation}
Of course, in order to predict $k^D_Q$, it is necessary to guess the expression for $\tau^\textrm{nl}_Q(\kappa)$. For instance, if a scaling law can be assumed, $\tau^\textrm{nl}_Q(\kappa)=A_Q\, \kappa^{-\alpha_Q}$, the dissipation wave number is given by:
\begin{equation}
k^D_Q\propto\left(\frac{1}{\nu\,A_Q}\right)^{1/(2-\alpha_Q)}.
\label{kdiss_time}
\end{equation}

Another typical length scale can be introduced \textit{via} Eq.\re{GB1} and under the assumption that a stationary state can be reached:
\begin{equation}
\Pi_Q^<(\kappa)=s_Q - 2\, \nu \, \int_0^\kappa\, dp\, p^2\, Q(p). \label{GenFlux}
\end{equation}
This expression can be used to obtain an estimate of the viscous scale $k_Q^\nu$ at which the viscous term becomes important when compared to $s_Q$ by assuming that the spectrum $Q(p)$ follows a power law: $Q(\kappa)=B_Q\, \kappa^{-\beta_Q}$:
\begin{equation}
2\, \nu\, \int_0^{k_Q^\nu}\,dp\, p^2\, B_Q\, p^{-\beta_Q} \approx s_Q,
\label{diss_estimate}
\end{equation}
which leads to
\begin{equation}
k_Q^\nu\propto \left(\frac{s_Q}{\nu\, B_Q}\right)^{1/(3-\beta_Q)}.
\label{kdiss_diss}
\end{equation}
For a conserved quantity, the spectrum has to fall off for $\kappa>k_Q^\nu$ otherwise the dissipation would exceed the injection rate and consequently the nonlinear term must vanish. It is thus expected that $k_Q^\nu=k_Q^D$. However, for a non conserved quantity, the dissipation may exceed the injection rate since the nonlinear term does not necessarily vanish. Thus, nothing prevents the spectrum to remain $Q(\kappa)=B_Q\, \kappa^{-\beta_Q}$ after $k_Q^\nu$ and $k_Q^\nu$ may be smaller than $k_Q^D$.


\section{Energy and helicity dissipation scales}

We first consider the cascade of energy. The total energy injection rate is then usually noted $s_E=\epsilon$ and the Kolmogorov energy spectrum can be derived,
\begin{equation}
E(k)=C_E\, \epsilon^{2/3}\,k^{-5/3}\,,
\label{Kol_E}
\end{equation}
in the inertial range. The estimate for the dissipation wavenumber based on the equality between the characteristic time scales requires an expression for $\tau^\textrm{nl}_E(\kappa)$. Various proposals can be found in the literature, but all yield the same scaling since they are built with only $\kappa$ and $\epsilon$, assuming the viscosity does not influence the non-linear characteristic time:
\begin{equation}
\tau^\textrm{nl}_E(\kappa)\propto \kappa^{-2/3} \epsilon^{-1/3}
\end{equation}
which means $A_E\propto \epsilon^{-1/3}$ and $\alpha_E=2/3$. Consequently, expression~(\ref{kdiss_time}) yields
\begin{equation}
k_E^D\propto \left(\frac{1}{\nu\, \epsilon^{-1/3}}\right)^{3/4}\propto\left(\frac{\epsilon}{\nu^3}\right)^{1/4}. \label{kE}
\end{equation}
Similarly, the Kolmogorov spectrum implies $\beta_E=5/3$ and $B_E\propto\epsilon^{2/3}$, and the expression~(\ref{kdiss_diss}) yields the same estimate:
\begin{equation}
k_E^\nu\propto \left(\frac{\epsilon}{\nu\, \epsilon^{2/3}}\right)^{3/4}\propto\left(\frac{\epsilon}{\nu^3}\right)^{1/4}.
\end{equation}

We now consider the helicity cascade. Both studies presented in \cite{Ditlevsen01a} and \cite{Chen03} make the assumption that the characteristic time of non-linear transfer of energy and helicity are the same: $\tau^\textrm{nl}_H(\kappa)=\tau^\textrm{nl}_E(\kappa)$. Since, both energy and helicity are dissipated by linear viscous processes, their dissipation characteristic time are obviously identical $\tau^\textrm{diss}_H(\kappa)=\tau^\textrm{diss}_E(\kappa)=1/(\nu\, k^2)$. In that case, the dissipation wavenumber for energy and helicity obtained by comparing the non-linear transfer time to the dissipation time must coincide:
\begin{equation}
k_H^D\propto k_E^D\propto\left(\frac{\epsilon}{\nu^3}\right)^{1/4}. \label{kHD-Scaling}
\end{equation}
Also, the equality of the non-linear transfer time is also known to imply the following helicity spectrum :
\begin{equation}
H(k)= C_H \delta\, \epsilon^{-1/3}\, k^{-5/3}\,,
\label{Kol_H}
\end{equation}
where $C_H$ is a dimensionless constant and $\delta$ is the helicity injection rate. In that case, the formula~(\ref{kdiss_diss}) with $s_H=\delta$ and $B_H=\delta \epsilon^{-1/3}$ leads to the same expression
\begin{equation}
k_H^\nu\propto\left(\frac{\delta}{\nu\, \delta\, \epsilon^{-1/3}}\right)^{3/4}\propto\left(\frac{\epsilon}{\nu^3}\right)^{1/4}.
\end{equation}
Hence, both approaches yields the same result and tend to confirm the equality between the helicity and the energy dissipation scales. However, although the equality of both dissipation scales is so obvious, the analysis becomes a bit more involved when using the helical decomposition of the energy and helicity spectra.


\section{Helical decomposition of spectra}

Following the approach presented in \cite{Ditlevsen01a}, the Fourier modes of both the velocity and the vorticity are expanded using a basis of polarised helical waves $\textbf{h}^{\pm}$ defined by
$i \textbf{k} \times \textbf{h}^{\pm} = \pm k \textbf{h}^{\pm}$~\cite{Waleffe92}:
\begin{align}
\textbf{u}(\textbf{k})&=u^+(\textbf{k})\, \textbf{h}^+ +u^-(\textbf{k})\, \textbf{h}^-,\\
\boldsymbol{\omega}(\textbf{k})&=k\, u^+(\textbf{k})\, \textbf{h}^+ -k\, u^-(\textbf{k})\, \textbf{h}^-.
\end{align}
The energy and helicity carried on by the mode $\textbf{u}(\textbf{k})$ respectively become
\begin{eqnarray}
	\textbf{u}(\textbf{k}) \cdot \textbf{u}^*(\textbf{k})/2&=&(|u^+(\textbf{k})|^2 + |u^-(\textbf{k})|^2)/2, \quad \quad \quad \label{Ewal}\\
  \textbf{u}(\textbf{k}) \cdot \boldsymbol{\omega}^*(\textbf{k})/2&=&k(|u^+(\textbf{k})|^2 - |u^-(\textbf{k})|^2)/2. \quad \quad \quad \label{Bwal}
\end{eqnarray}

Isotropy is again assumed and both the energy $E(k)$ and the helicity $H(k)$ spectra are considered to be functions of $k=|\textbf{k}|$. Introducing the spectral densities of energy and helicity for the helical modes ($\pm$) yields:
\begin{eqnarray}
	E(k)&=&E^+(k)+E^-(k)\,, \label{E(k)} \\
	H(k)&=&H^+(k)+H^-(k)=k[E^+(k)-E^-(k)]\,. \label{H(k)}
\end{eqnarray}
Their equations of evolution have exactly the structure~(\ref{general_balance}). Moreover, all these quantities are dissipated through viscous effect and their linear dissipation time scale is again $\tau^\textrm{diss}(\kappa)=1/(\nu\, k^2)$. Guessing their non-linear characteristic time is however much more difficult. Indeed, non-linear transfers can transform $E^{<+}(\kappa)$ not only in $E^{>+}(\kappa)$ but also in $E^{<-}(\kappa)$ and $E^{>-}(\kappa)$. Moreover, $E^+$ and $E^-$ are not separately conserved by the non-linear terms. Hence, invoking the equality of characteristic time scales to estimate the dissipation scales of these quantities is not really possible.

It is however quite easy to estimate their spectra from~(\ref{Kol_E}) and~(\ref{Kol_H}):
\begin{subequations}\label{E+-}
\begin{eqnarray}
	E^+(k)= \frac{C_E}{2} \varepsilon^{2/3}k^{-5/3} +  \frac{C_H}{2} (\delta / \varepsilon^{1/3})k^{-8/3}, \label{E+}\\
	E^-(k)=  \frac{C_E}{2} \varepsilon^{2/3}k^{-5/3} -  \frac{C_H}{2} (\delta /\varepsilon^{1/3})k^{-8/3},\label{E-}
\end{eqnarray}
\end{subequations}
which are the equations (9) and (10) of \cite{Ditlevsen01a}. As a consequence, the leading order in $k$ must be given by
\begin{eqnarray}
E^{\pm}(k) = \frac{C_E}{2} \varepsilon^{2/3}k^{-5/3}, \; H^{\pm}(k) =\pm \frac{C_E}{2} \varepsilon^{2/3}k^{-2/3}.
\label{EH+-}
\end{eqnarray}

By construction, the range of validity of\re{E+-} and\re{EH+-} is the same as that of the scaling laws\re{Kol_E} and\re{Kol_H} of $E(k)$ and $H(k)$. It is therefore bounded by $k_E^D=k_H^D$.

On the other hand, formula~(\ref{kdiss_diss}) yields an estimate of the scale from which on the dissipative term must be considered in the evolution of $E^\pm$ and $H^\pm$. It leads to
\begin{eqnarray}
k_{E^\pm}^{\nu}&\propto&\left(\frac{\epsilon}{\nu^3}\right)^{1/4}; \\
\textrm{   and   } k_{H^\pm}^{\nu}&\propto&\left(\frac{\delta}{\nu \epsilon^{2/3}}\right)^{3/7}
\propto\left(\frac{\delta^{3}}{\nu^3\epsilon^2}\right)^{1/7}.
\end{eqnarray}

As noted by Ditlevsen and Giuliani~\cite{Ditlevsen01a}, $k_{H^\pm}^{\nu}<k_{E^\pm}^{\nu}=k_E^{D}$. Indeed, the helicity injection rate is at most $k_F\, \epsilon$, so that
\begin{equation}
k_{H^\pm}^{\nu}< \left(\frac{k_F^3 \epsilon}{\nu^3}\right)^{1/7}= \left(\frac{k_F}{k_E^{D}}\right)^{3/7}\ k_E^{D},
\end{equation}
and since $k_F<k_E^{D}$ in the turbulent regime, $k_{H^\pm}^{\nu}<k_E^{D}$. However, there is no reason to identify the helicity dissipation scale as $k_{H^\pm}^{\nu}$. Clearly, the spectrum of $H^+(k)$ and $H^-(k)$ can not deviate from  the scaling $k^{-2/3}$ and fall off in the range $k_{H^\pm}^{\nu}<k<k_E^{D}$. Indeed, if these quantities decay faster that $k^{-2/3}$ after $k_{H^\pm}^{D}$, then the quantities $E^\pm(k)=\pm H^\pm(k)/k$ will decay faster than $k^{-5/3}$ and $k_{H^\pm}^{\nu}$ would be identified as the end of the inertial range which is known to actually extend down to $k_E^D$. Considering the Eq.~(\ref{GenFlux}) applied to $Q=H^\pm$ allows to better understand the meaning of $k_{H^\pm}^{\nu}$:
\begin{eqnarray}
{\Pi_H^{\pm<}} &=& \delta^{\pm} \mp \frac{3}{7}C_E \nu \varepsilon^{2/3}k^{7/3}\label{PiH+-}.
\end{eqnarray}
For low values of $k\ll k_{H^\pm}^{\nu}$, the fluxes are constant and equal to $\delta^\pm$. However, for $k\gg k_{H^\pm}^{\nu}$, the dissipation of $H^\pm$ is stronger than the injection rate $\delta^\pm$ and the nonlinear flux has to scale like $k^{7/3}$. Ditlevsen and Giuliani referred to this scale the dissipation scale of $Q^\pm$. However, as argued above, this does not correspond to the end of the helicity spectrum.

\section{Helical shell model analysis\label{sec-NSHelAn}}

Shell models are built to describe the exchange of physically relevant quantities between the various scales of a turbulent flow. The Fourier space is divided into a set of shells which are logaritmically spaced. A field (like velocity for instance) is represented by very few (1 or 2) complex variables in each shell. These models allow to investigate turbulence properties at a much lower numerical cost than direct numerical simulation. Actually, the resolutions achievable in DNS are still too limited to distinguish clearly $k_{H^\pm}^{\nu}$ from $k_{E}^{D}$~\cite{Chen03}. On the other hand, as will be done below, Reynolds numbers - defined at scale $k_F$ as $Re=\varepsilon^{1/3} k_F^{-4/3} \nu^{-1}$ - as large as $10^7$ can be reached with a shell model.

 Helical shell models were developed in~\cite{Benzi96}, and
are based on the helical decomposition of Fourier modes~\cite{Waleffe92}. Two dynamical variables are used per shell to represent both the helical components of the velocity field. As pointed out in~\cite{TLFPDC}, such helical models can be retrieved from the helical triadic systems of the Navier-Stokes equations in helical basis. Four simple models can be expressed in a single formula:
\begin{subequations}\label{Model}
\begin{equation}
d_t u_n^\pm = W_n^\pm -\nu k_n^2 u_n^\pm + f_n^\pm,
\end{equation}
with
\begin{eqnarray}
W_n^\pm &=& i k_n \big[ (s_1\la - s_2 \la^2 ) u_{n+1}^{\pm s_1} u_{n+2}^{\pm s_2}\nonumber\\
&+& (s_2 \la -\la^{-1})\ u_{n-1}^{\pm\ s_1} u_{n+1}^{\pm s_2 s_1} \\
&+&(\la^{-2} - s_1 \la^{-1})\ u_{n-2}^{\pm s_2} u_{n-1}^{\pm s_1 s_2}\big]^*,\nonumber
\label{ModelW}
\end{eqnarray}
\end{subequations}
where each model is obtained for one particular choice of $(s_1, s_2)$ with $s_1,\ s_2 =\pm 1$.
In (\ref{Model}) the parameter $\lambda$ is the logarithmic shell spacing and the wave number is defined as $k_n=k_0 \lambda^n$.

In the absence of forcing and viscosity $\nu$, the shell model (\ref{Model}) conserves total energy $E$ and helicity $H$~\cite{TLFPDC}.
\begin{eqnarray}
E= \sum_{n=1}^N E_n, \quad \quad
H= \sum_{n=1}^N H_n,
\end{eqnarray}
where $N$ is the number of shells in the model. The energy $E_n$ and helicity $H_n$ in shell $n$ are defined as
\begin{eqnarray}
	E_n&= E_n^+ + E_n^-, \quad E_n^\pm&=\frac{1}{2}|u_n^\pm|^2 , \\
	H_n&= H_n^+ + H_n^-, \quad H_n^\pm&=\pm\frac{1}{2}k_n |u_n^\pm|^2.
\end{eqnarray}

Within the model, the fluxes of energy and helicity are defined as
\begin{eqnarray}
	\Pi_E^<(n) &=& \Pi_E^{+<}(n) + \Pi_E^{-<}(n),\\
	\Pi_H^<(n) &=& \Pi_H^{+<}(n) + \Pi_H^{-<}(n),
\end{eqnarray}
with the following explicit expressions:
\begin{eqnarray}
		\Pi_E^{\pm<}(n) &=& -\left . \left ( d_{t} \sum^{n}_{m=1}\frac 1 2 |u_m^\pm|^2 \right)\right |_{\textrm{NL}}\nonumber\\
		&&\qquad =  -\sum^{n}_{m=1}W_m^\pm u_m^{\pm*} + cc,\\
\Pi_H^{\pm<}(n) &=&-\left . \left ( d_{t} \sum^{n}_{m=1}\frac 1 2 (\pm k_m) |u_m^\pm|^2 \right) \right |_{\textrm{NL}} \nonumber \\
	&&\qquad= \mp \sum^{n}_{m=1}k_mW_m^\pm u_m^{\pm*} + cc,
\end{eqnarray}
where $\Pi^<_Q(k)$ is the flux (due to the non-linear term) of the quantity $Q$ leaving the region of wave numbers lower than $k$ and $\Pi^{\pm<}_{Q}(k)$ is the flux leaving either the `$+$' or the `$-$' variables of wave numbers lower than $k$.


The GOY model used in \cite{Ditlevsen01b} corresponds to $(s_1,s_2)=(-, +)$ in the helical picture~(\ref{Model}). In this case, two uncoupled sets of variables appear, namely: $(u_1^+, u_2^-, u_3^+ ,\dots)$ and $(u_1^-, u_2^+, u_3^- ,\dots)$. In the original version of the GOY model, only one of these sets is considered. Hence, in each shell $n$, the helicity is evaluated alternatively by $H^+_n$ or $H^-_n$, depending whether $n$ is odd or even. The cancellation of the leading terms in equation~(\ref{H(k)}) with the scaling~(\ref{E+}) and (\ref{E-}) does not occur. Therefore $H(k)$ can not be straightforwardly obtained with a GOY model. The fluxes presented in~\cite{Ditlevsen01b} are hence closer to $\Pi_H^{\pm<}(\kappa)$ than to $\Pi_H^<(\kappa)$ although, \textit{stricto sensu} they are neither of them.

On the other hand, the developments proposed in~\cite{Chen03} were illustrated by the SABRA version of the model corresponding to $(s_1,s_2)= (+,-)$ in which all variables are coupled. Both $H^+_n$ and $H^-_n$ are available within each shell $n$ and so is the total helicity $H_n$. In the following section, the work of~\cite{Chen03} is pursued and the energy and helicity spectra and fluxes are investigated.

\section{Numerical results}

The computation of the averaged helicity spectra, which is the difference of its two helical components and requires the cancelling of the leading terms demands very fine time stepping. Furthermore, very long simulations are required in order to obtain enough statistics. This is probably the reason why helicity spectra have not been reported so far~\cite{Ditlevsen01a,Ditlevsen01b,Chen03}, with the notable exception of~\cite{Stepanov09}. Very long and accurate integration of the shell model\re{Model} with $(s_1,s_2)= (+,-)$ have been performed. In these simulations, the forcing is concentrated on one single shell (the fourth) and provides constant energy and helicity injection rates. The rate of energy injection within the `$\pm$' variables is denoted $\varepsilon^\pm$. The rate of helicity injection is therefore $\delta^\pm=\pm\, k_F \varepsilon^\pm$. The phases are randomly chosen at each time step.

\begin{figure}
\begin{tabular}{@{}c@{\hspace{0mm}}c@{}}
    \epsfig{file=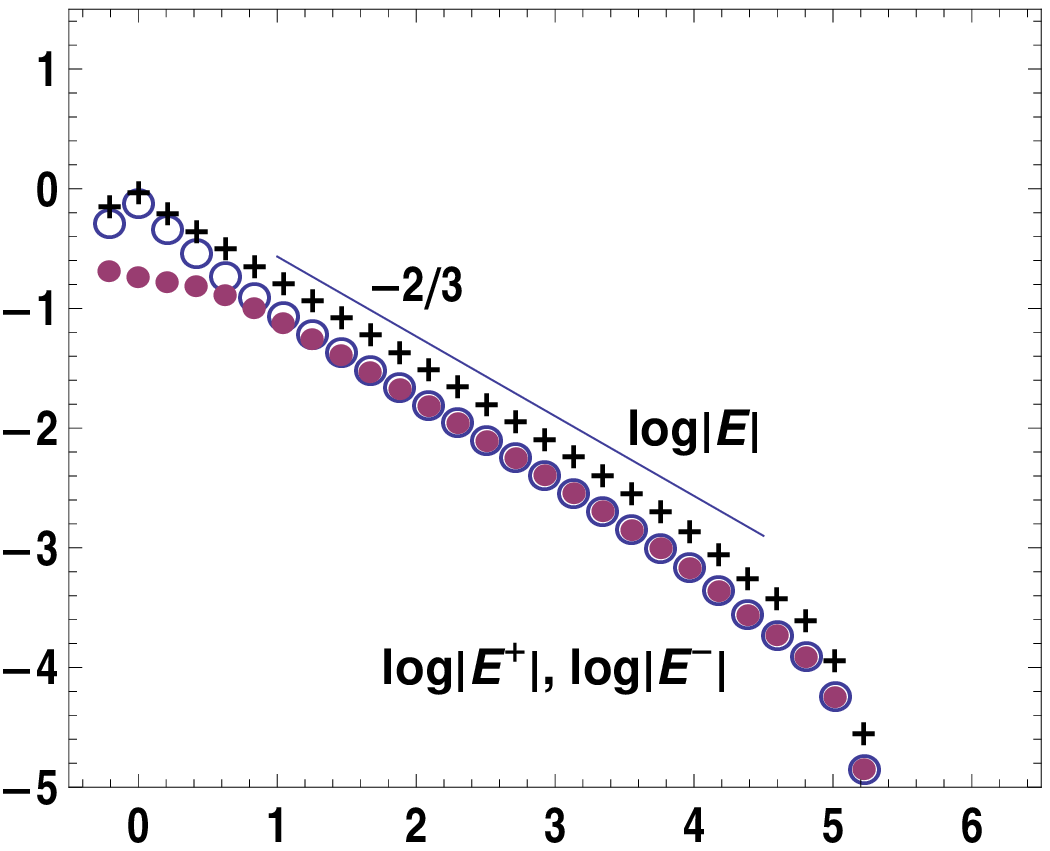,width=0.21\textwidth}
    &
    \epsfig{file=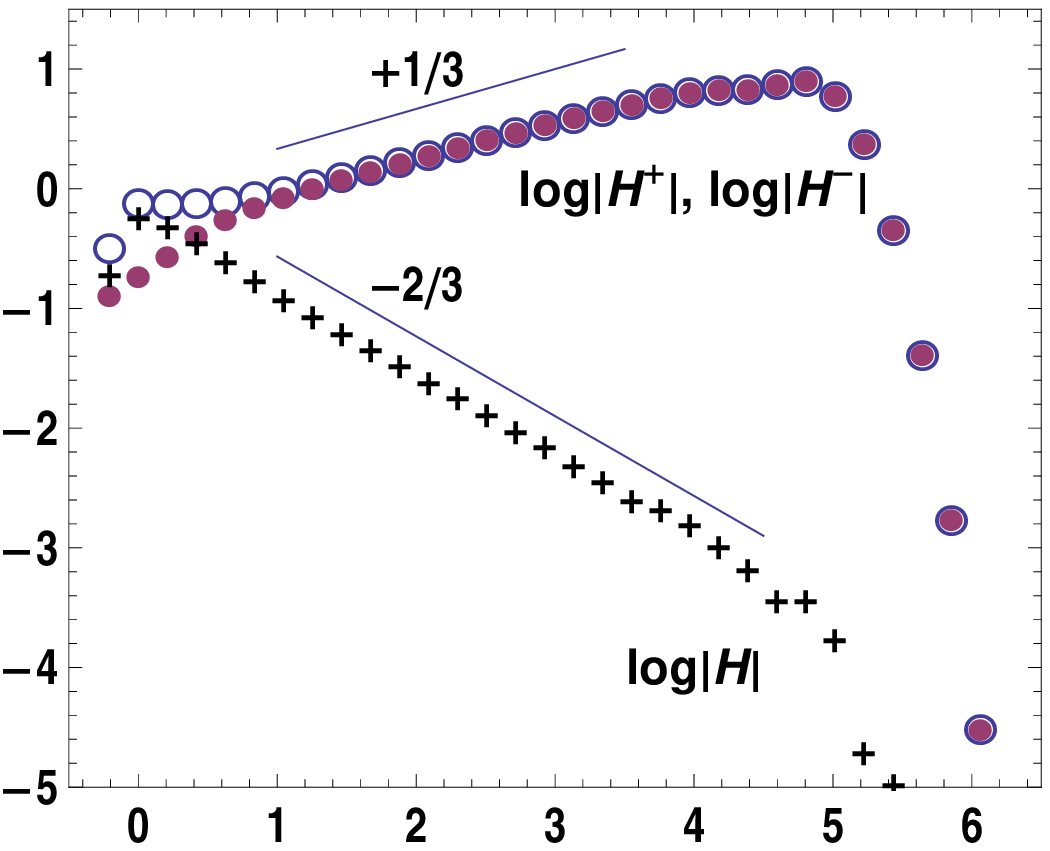,width=0.21\textwidth}
    \\*[-0.cm]
    \epsfig{file=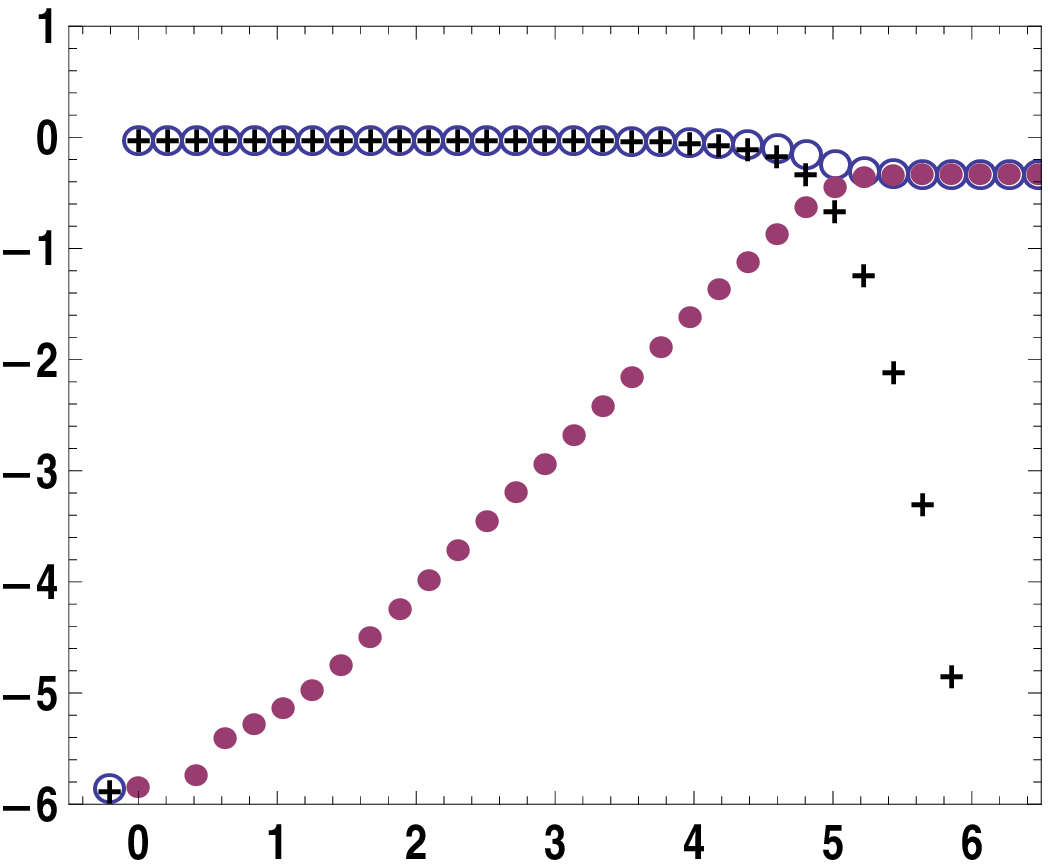,width=0.21\textwidth}
    &
    \epsfig{file=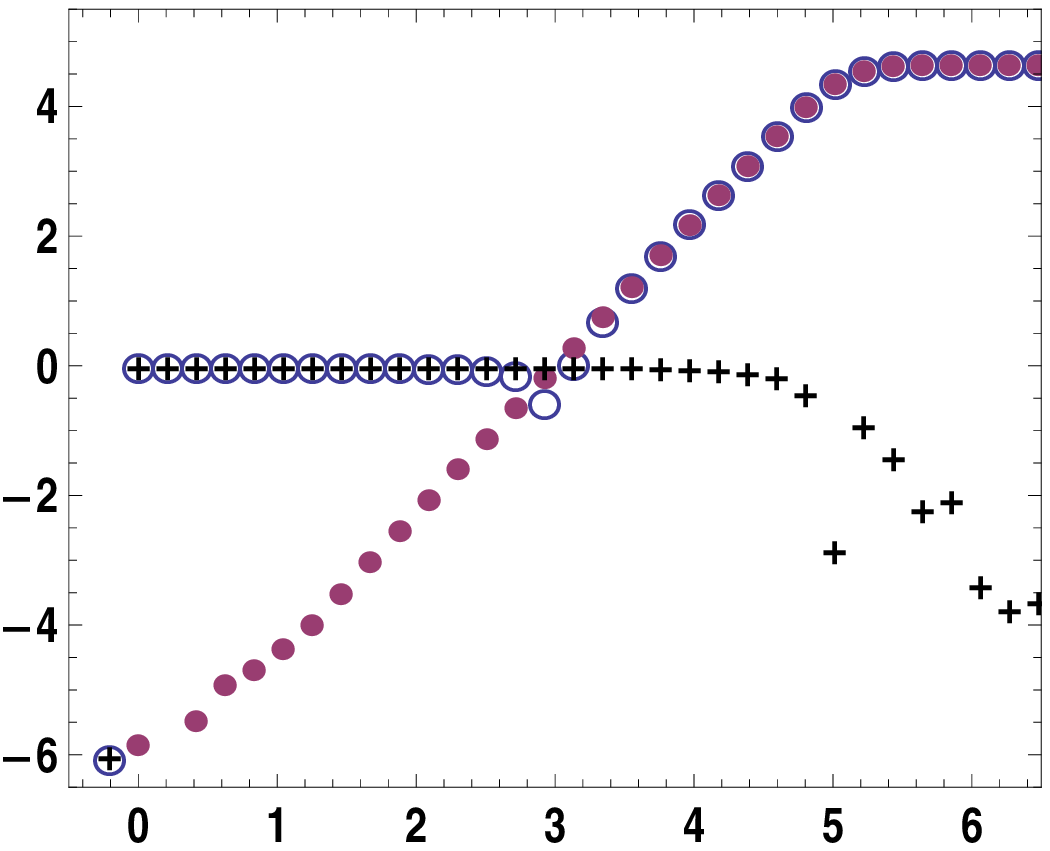,width=0.21\textwidth}
  \end{tabular}
\caption{{\it Helical case: $\varepsilon=\varepsilon^+=1$, $\delta^-=\varepsilon^-=0$. Energy and helicity plots are respectively represented on the left and right columns versus $\log k$.
The positive and negative helical modes are denoted by $\circ$ and $\bullet$, the sum of both modes by +.
The spectra (resp. fluxes) are represented in the top (resp. down) row.  \label{fighelical}}}

\end{figure}

\begin{figure}
\begin{tabular}{@{}c@{\hspace{0mm}}c@{}}
    \epsfig{file=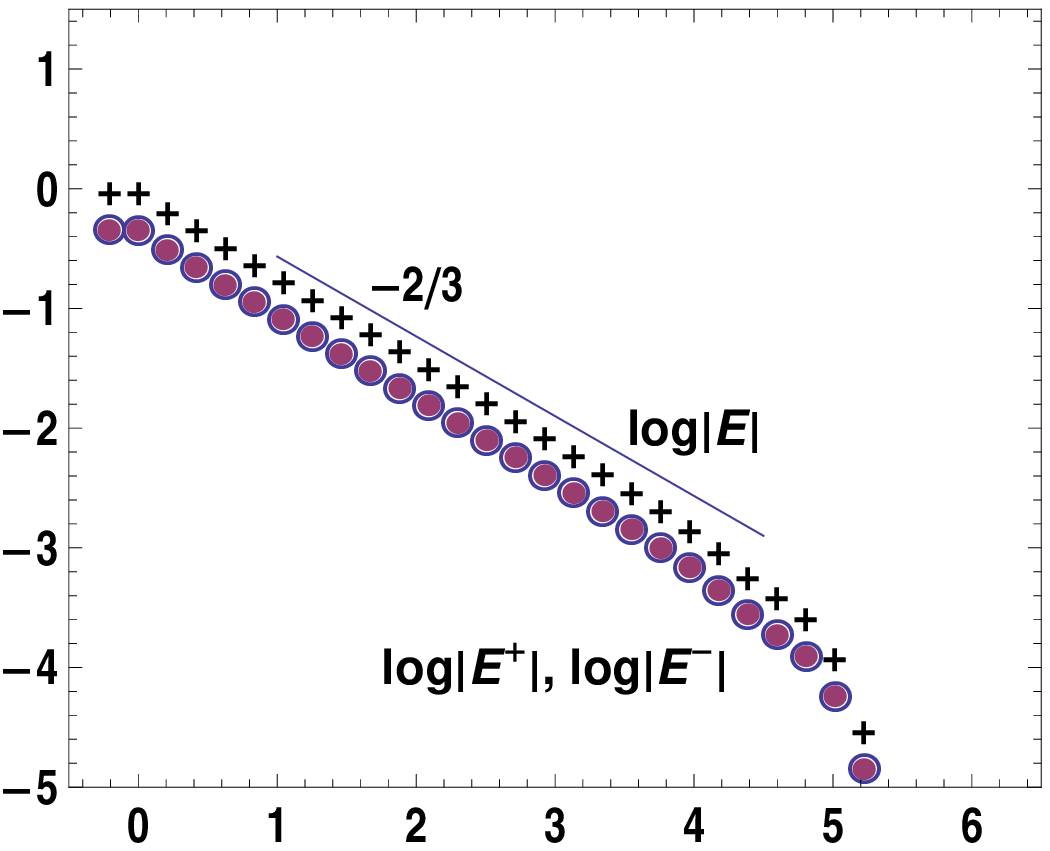,width=0.21\textwidth}
    &
    \epsfig{file=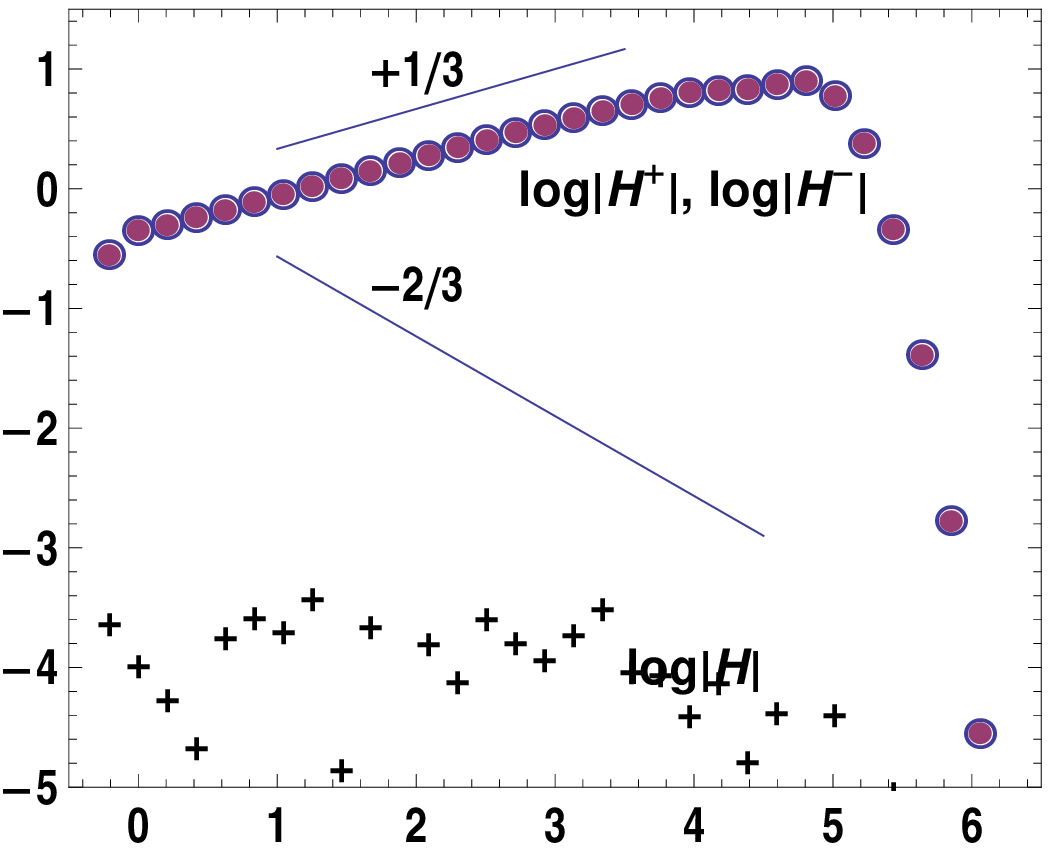,width=0.21\textwidth}
    \\*[-0.cm]
    \\*[-0.cm]
    \epsfig{file=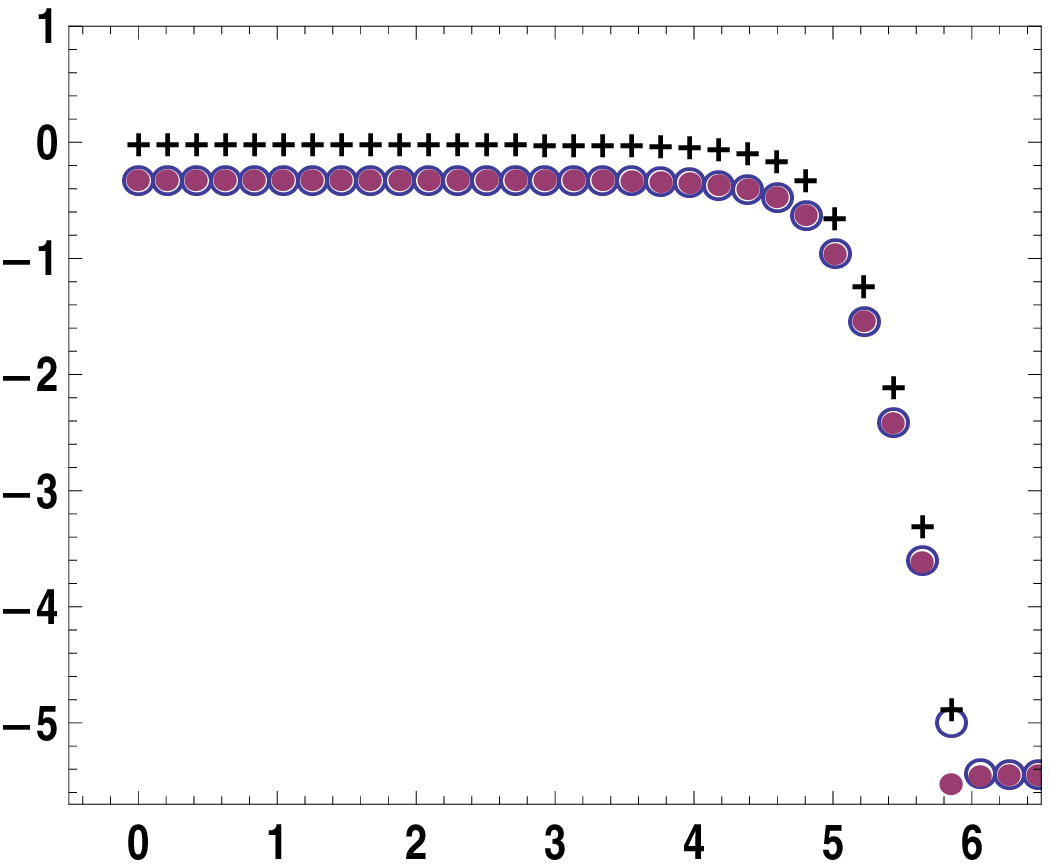,width=0.21\textwidth}
    &
    \epsfig{file=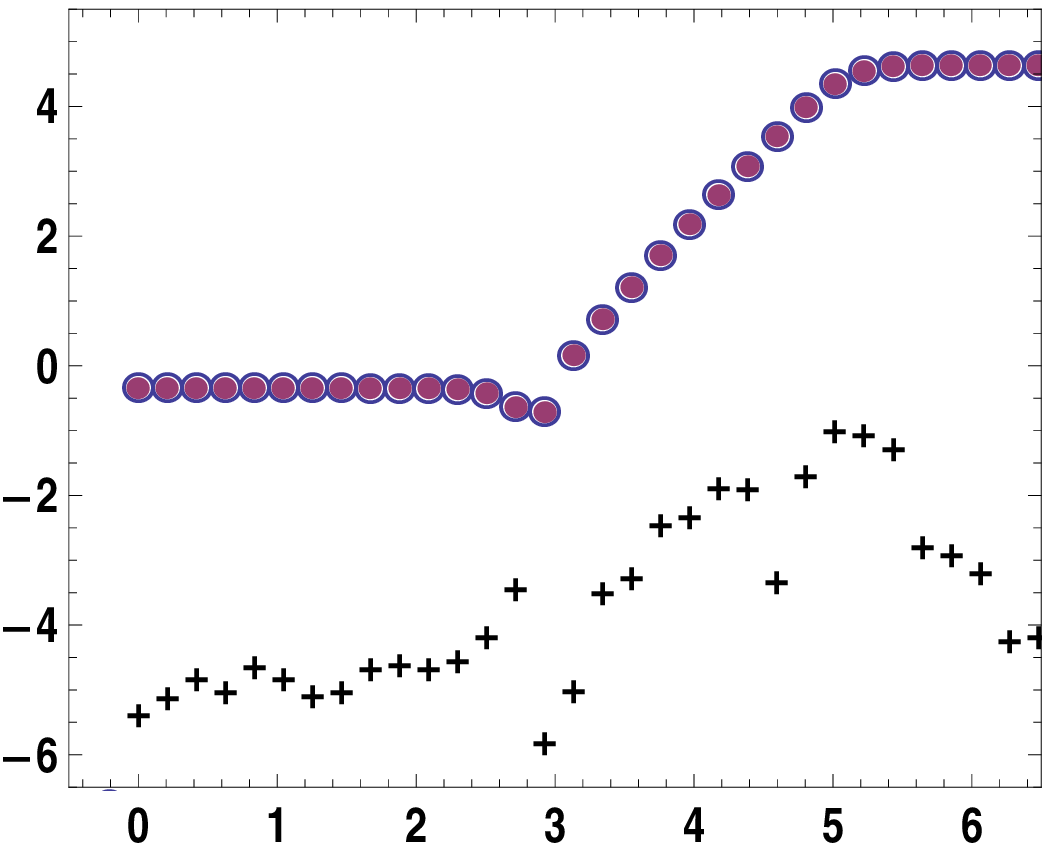,width=0.21\textwidth}
  \end{tabular}
\caption{{\it Same as figure \ref{fighelical} for the non helical case $\varepsilon^-=\varepsilon^+=1/2$, $\delta=0$.  \label{fignonhelical}}}
\end{figure}

In figures \ref{fighelical} and \ref{fignonhelical} the results are presented for
respectively a helical and a non helical case.
The parameters are $\nu=10^{-7}$ and $\ld=(1+\sqrt 5)/2$. The shell are labelled from $-2$ to $37$ with $k_n= \ld^n$. The total number of shells is thus $N=40$ and the forcing is concentrated in the third shell so that $k_F=1$.
For the helical case $\varepsilon^+=\varepsilon=1$, implying $\delta=\delta^+ =1$ and $\varepsilon^-=-\delta^-/k_F=0$. For the non helical case $\varepsilon^+=\varepsilon^-=1/2$, implying $\varepsilon=1$, $\delta^+ =- \delta^- =1/2$ and $\delta=0$.
In each figure, the left and right columns correspond respectively to energies and helicities.


The spectra are plotted in $\log$-$\log$ frames (upper row). Energies $E(k_n)$ and $E^{\pm}(k_n)$ scale in $k_n^{-2/3}$ corresponding to power spectral densities in $k^{-5/3}$ in agreement with (\ref{Kol_E}) and (\ref{EH+-}). Helicities $H^{\pm}(k_n)$ scale in $k_n^{1/3}$ corresponding to power spectral densities in $k^{-2/3}$ in agreement with (\ref{EH+-}).
In the helical case, the total helicity $H(k_n)$ scales in $k_n^{-2/3}$ corresponding to a power spectral density in $k^{-5/3}$ in agreement with (\ref{Kol_H}). In the non helical case $H(k_n)$ is the sum of two opposite quantities $H^{\pm}(k)$ and has no clear
scaling. Compared to $H^{\pm}(k_n)$ it can be considered as negligible, in agreement with (\ref{Kol_H}) taking $\delta=0$. Note that all spectra manifestly extend up to the Kolmogorov scale $k_E^D\sim10^5$

The nonlinear fluxes are plotted in $\log$-$\log$ frames (lower row). For the helical case, the total energy flux as well as the energy flux of $E^+$ are constant and dominated by $\epsilon=1$ up to the Kolmogorov scale. On the contrary the flux of $E^-$ has no component corresponding to the injection since $\epsilon^-=0$ so that its spectrum is dominated for low $k$ by the viscous term and is proportional to $k^{4/3}$. The viscous scale $k_{H^\pm}^\nu$ is clearly identified on the helicity flux for $H^+$. For $k<k_{H^\pm}^\nu$, the flux is constant and dominated by $\delta^+$ while for $k<k_{H^\pm}^\nu$ the injection is sub-leading and the flux scales like  $k^{7/3}$. Remarkably, the viscous scale $k<k_{H^\pm}^\nu$ is also very clearly observed even in the non-helical case.

\section{Conclusion}

The present study has allowed to identify two different length scales related to to the dissipation of a quadratic quantity $Q$ in Navier-Stokes turbulence. The first one is the traditional dissipation scale that marks the end of the power law in the spectrum of $Q$ due to the dominant effect of the viscosity. The second scale, referred to as the viscous scale, corresponds to the beginning of the range in which viscous effect have to be taken into account. Clearly, for the kinetic energy, the viscous and the dissipation scales coincide. However, for non conserved quantities, such as the positive and negative part of the helicity, these two scales are different. Although the viscous scale cannot be measured from the spectra, it is easily identified from the nonlinear fluxes. This has been shown using shell models.

This approach reconcile the analysis of~\cite{Ditlevsen01a,Ditlevsen01b} and~\cite{Chen03}. Strictly speaking, the scale $k_{H^\pm}^\nu$ cannot be interpreted as the dissipation scale for helicity. Both direct shell model integration and helical components analysis show that the helicity cascade develops down to the Kolmogorov scale. However, this scale is indeed relevant in the analysis of the nonlinear flux of helicity and plays a role even when the flow is globally non helical.

Beyond the issue of helicity dissipation scale which is now clarified, this study stresses how much caution is required when studying the effect of helicity on turbulence dynamics with a GOY model~\cite{Bowman06}. Other models like the one used here or those presented in~\cite{TLFPDC} or~\cite{Stepanov09} are highly preferable.

{\bf Acknowledgements.} F.P. and R.S. are grateful to Peter Ditlevsen for useful discussions. This work has been
supported by the contract of association EURATOM - Belgian state. The content of the
publication is the sole responsibility of the authors and it does not necessarily represent
the views of the Commission or its services. D.C. and T.L. are supported by the Fonds
de la Recherche Scientifique (Belgium). The support of the parallel computations on the
supercomputer SKIF MSU "Tchebyshoff" (under project 09-P-1-1002) is kindly appreciated.

\end{document}